\documentclass[letterpaper,12pt,final]{article}
\usepackage[margin=2.5cm]{geometry}
\usepackage[numbers,square]{natbib}
\usepackage{url}
\usepackage{setspace}
\usepackage{graphicx}

\date{}

\begin{document}

\title{Looking before Leaping: Creating a Software Registry}
\author{Alice Allen \\ Astrophysics Source Code Library, aallen@ascl.net \\ Judy Schmidt \\ Astrophysics Source Code Library}

\maketitle

\begin{abstract}

What lessons can be learned from examining numerous efforts to create a repository or directory of scientist-written software for a discipline? Astronomy has seen a number of efforts to build such a resource, one of which is the Astrophysics Source Code Library (ASCL). The ASCL (ascl.net) was founded in 1999, had a period of dormancy, and was restarted in 2010. When taking over responsibility for the ASCL in 2010, the new editor sought to answer the opening question, hoping this would better inform the work to be done. We also provide specific steps the ASCL is taking to try to improve code sharing and discovery in astronomy and share recent improvements to the resource.

\end{abstract}

\section{Introduction}

The Astrophysics Source Code Library (ASCL) was founded in 1999 to improve the transparency and reproducibility of research by making the software used in astrophysics research discoverable for examination \citep{RJN_aas1999}. We also encourage reuse of software and advocate for accurate software citation; citing codes increases transparency and gives software authors a way to demonstrate the value of their work, thus providing incentive to make their programs available. Software methods are vital to research \citep{2013Sci...340..814J, 2012Natur.482..485I, 2011Sci...334.1226P}; the importance of computational methods in astronomy is underscored by the existence of journals devoted specifically to these methods \citep{2011ASPC..442..655G}.

For the ASCL, sustainability is not nearly so ambitious as that defined in {\em The Blind Men and the Elephant: Towards an Empirical Evaluation Framework for Software Sustainability} from the 2013 Workshop on Sustainable Software for Science: Practice and Experiences (WSSSPE): ``a measure of a systems extensibility, interoperability, maintainability, portability, reusability, scalability, and usability'' \citep{JORSjors.ao}. Our sustainability model is much smaller: {\em Is the source code available to be read, and can it be found easily?} Preservation and discovery are the most important attributes -- if the code is not preserved and cannot be found, no other attributes of sustainability matter!

Several libraries or registries of astrophysics codes have been created over the past two decades. Each endeavor offered opportunities for researchers to share their software techniques and improve the reproducibility of research, but none reached the tipping point of widespread adoption by the community. What lessons can be learned from examining these efforts?

\section{Looking Back and Around}

Though there were online sites for individual codes and a few other smaller collections, perhaps the first robust online resource for astronomy software was the Astronomical Software Directory Service (ASDS), conceived in 1993 by Robert J. Hanisch, Harry E. Payne, and Jeffrey J.E. Hayes \citep{RJH_adass1994}. Funded by the NASA Astrophysics Data Program \citep{HEP_aas1998}, ASDS was not a repository or library; rather, it maintained information about codes, but the codes and related files such as documentation were stored on other sites, typically personal or institutional websites. This resource eventually provided information on 56 programs. In 1998, ASDS changed its name to the Astronomical Software and Documentation Service as it expanded to include telescope and other instrument manuals; the service ended about 2000. The ASDS had expected that ``software providers would be their own metadata managers. But they didn't do it.'' Over time,``the directory service started to be out of date'' \citep{Han..private..2011}, making it increasingly less useful. 

In 1999, Robert Nemiroff at Michigan Technological University and John F. Wallin, then of George Mason University and now at Middle Tennessee State Univeristy, founded the online Astrophysics Source Code Library (ASCL) to house programs of use to the community \citep{RJN_aas1999} and serve as a download site for them. This was a volunteer, spare-time endeavor; between 1999 and 2002, approximately 40 codes were deposited. Nemiroff said that ``Getting people to post their codes in ASCL was like pulling teeth. Nobody submitted anything. We had to go out there and get them ourselves.'' \citep{Nem..whois..} In 2003, a new volunteer editor-in-chief was sought for ASCL; none was forthcoming, and with developments elsewhere to provide a code repository, the ASCL remained available but work on it ceased.

In 2003, Michael Remijan and Robert Brunner from the National Center for Supercomputing Applications (NCSA) at the University of Illinois announced Astroforge, which used the SourceForge model for open source software but focused on the needs of astronomers \citep{Remijian..2003}. Development of the resource was a three-year project  \citep{Bru..AISRP}. The site, which was at http://www.astroforge.net, no longer exists; it was discontinued due to lack of funding \citep{Bru..private..2011}.

We know of four other code directories, though certainly there are many project groups and individuals who pull together such information for their own work, team, or subspecialty. One such subspecialty repository is Astr{\it O}matic\footnote{http://www.astromatic.net/} for astronomical pipeline software; another is the codes wiki for computational fluid dynamics\footnote{http://www.cfd-online.com/Wiki/Codes}. 

Carlo Baffa, Elisabetta Giani, and Alessio Checcucci of the Istituto Nazionale di Astrofisica (INAF) created SkySoft\footnote{http://www.skysoft.org/html/index.php} in 2003 as ``Yet Another Astronomical Software Directory, but with a different overall approach as a community-supported directory, to which everyone can contribute, both developers and end-users.'' \citep{2004ASPC..314..157B} The site has entries for 309 codes, many added in 2003 and 2004; since 2007, 18 software packages have been added. Software used in research as well as development tools and management codes are included, and the site provides space for news items, which are frequently updated, and links to astronomy institutions, resources, and other software sites.

The Astro-Code Wiki\footnote{http://www.astrosim.net/code/doku.php}, created by AstroSim - European Network for Computational Astrophysics\footnote{http://www.astrosim.net/index.html}, contains approximately 55 codes and has had sporadic updates. AstroSim was a five-year project ``to bring together European computational astrophysicists'' running from October, 2006 until September, 2011 \citep{Moore..astrosim}; its focus was on comparison of codes for suitability for specific tasks. Once the funding ran out, the project ended; the Astro-Code Wiki is still available, though the most recent code addition was made in 2013.

Another repository called Astro-Sim was established in April, 2007 by Steffen Brinkmann of the High Performance Computing Center Stuttgart (HLRS) of the University of Stuttgart \citep{Brinkmann..astro.sim}. This site, which housed about thirty codes, also provided forums for discussion and links to other tools and libraries. The site\footnote{http://astro-sim.org/, no longer working} had been updated as recently as August, 2010, but now appears to have been deleted. 

Keith Shortridge discussed the need for communication and sharing among astronomical software users and authors at the Astronomical Data Analysis Software and Systems (ADASS) conference in 2008; included in his talk was information about a TWiki site called AstroShare, created in October 2008, which was intended to permit information and codes sharing\citep{2009ASPC..411....3S}. The site housed about 30 packages and had areas for discussion of topics such as releasing software, social media, and middleware. A flurry of activity occurred shortly after Shortridge's presentation and subsequent publication of his talk but did not continue. In mid-2014, the site had 126 registered users and showed a last update of November 2012; since then, however, the URL for the site\footnote{https://www.astroshare.org/bin/view/AstroShare/WebHome, no longer working} stopped working and as of this writing (August 2015), the domain is used by the Chilean Virtual Observatory.

\section{Learning From the Past}

What we learned from these endeavors was instrumental in our work to restart and grow the ASCL. Each of the efforts to aid communication and share knowledge of codes useful for astrophysics has offered valuable information to the community, yet a number of common factors inhibited the growth and use of these software directories.

The experience of ASDS demonstrates that metadata curation is an ongoing requirement for a resource to be useful and used, and that code authors do not want to take on that task \citep{Han..private..2011}. The original ASCL was not successful according to Nemiroff because ``most people just didn't know about it and had few ways to find it.'' He came to realize that consistent exposure is needed for a resource to become known and used \citep{Nem..private..2011}. SkySoft is also affected by a lack of marketing, replicating the ASCL's experience in its earliest days: an initial announcement, a splash or two that generated activity, and then a gradual fading of interest. 

Curation and marketing are not the most difficult hurdles; the greatest inhibitors relate to human nature, including the unwillingness of scientists to share their codes openly, the effect of the lack of an adequate reward system for software authorship, and the competitive environment in astronomy.

A depressingly small number of programmers are willing to share their work openly even on their own websites. Most emails that we have sent to software authors to request either a download site for or an archive file of a code we have found in a research paper go unanswered; authors who do reply usually say their codes are not available. The common reasons code authors are reluctant to share their work are covered more fully elsewhere \citep{barnes2010pcc, 2013A&C.....1...54S, JORSjors.bd}.

Nemiroff's assumption that software authors would be eager to deposit their code proved false \citep{Nem..whois..}. Indeed, experience shows that even authors with open code are reluctant to provide copies of their software to a repository site; many prefer to keep their work close to them, especially if they continue to refine and develop these codes. Most efforts have relied on software authors to submit codes, yet most authors are unwilling to submit information and maintain metadata stored in a software directory, especially as the payback for doing so may be uncertain. 

Astroforge ceased to exist when its initial funding ended \citep{Bru..private..2011}. The AstroSim project was funded for five years and is now closed; though its Astro-Code Wiki is still available, it is not updated often and is not actively advertised. ASDS, too, ended when its support did. Success of a software directory requires a change in community attitudes and behavior to one of routine code sharing. Absent outside pressures, change in how a discipline works often takes time, longer than a funding cycle of three or even five years. Without uptake by the community, a new software directory is unlikely to receive additional funding, and as the funding goes away, so too may the resource. 

\section{ASCL 2.0}

In 2010, the ASCL was restarted with the decision to move the existing code entries to a phpBB site (Starship Asterisk) that houses forums for discussing the Astronomy Picture of the Day (APOD)\footnote{http://apod.nasa.gov/; Nemiroff is a co-creator and co-editor of this very popular astronomy site.}, educational resources for astronomy, and other topics of interest to APOD viewers, and a new volunteer editor was found. The ASCL initially had been a code repository; as software may undergo additional sporadic or continuous updating, it was decided that linking to a download site and dropping the requirement of software deposit would be sufficient for discovery. Requirements for metadata that may prove ephemeral and that are typically found on the download website (or inherent in the software) were also dropped to increase the accuracy of each entry, as was code categorization; with Starship Asterisk offering full-text iterative searching, it was felt that feature offered more opportunity for discovery than static categorization.

Each code record was posted to a separate discussion thread in a new forum dedicated to the ASCL; threads on this forum were ordered alphabetically by the name of the software. Authors and other interested parties were welcome to post to these threads, to provide updates, ask questions, or share other information they wanted to post. In the four years the ASCL was housed on Starship Asterisk, few people took advantage of this ability, though some did subscribe to particular threads, or the forum itself, to be notified of any changes to those threads or the forum.

It was at this time the examination of other efforts described in section 2 took place. Examining the other projects and talking with those who had run them was instructive, and provided guidance in developing change management strategies to increase the likelihood the ASCL would become a useful and used resource for the community. With the goals of improving the discoverability of research software, increasing the transparency, integrity, and falsifiability of research, and perhaps even helping to improve the efficiency of the field by encouraging code reuse, we developed a plan to guide the future of the ASCL.

\section{Creating a Sharing Community}

Space does not allow a complete list of strategies that can be employed to inspire a community to new behaviors, but to start to move the astronomy community to one that routinely shares codes, we can:

\begin{enumerate}
\item ensure there is a way to share software (build an infrastructure) that
\item provides incentives for making codes available, and 
\item enlist/involve others with appropriate credibility in the community to endorse the effort, 
\item market the effort effectively, 
\item engage the community to learn what barriers and incentives exist and
\item mitigate these barriers and nurture the incentives, 
\item examine and reach out to other communities, and then finally,
\item be patient.
\end{enumerate}

Let's look at a few ways we implemented these steps when restarting the ASCL. 
\\
\\
\noindent {\em Build an infrastructure}
\\
In 2010, the existing codes were moved from the original ASCL HTML pages to the phpBB discussion forum that houses, among other things, the discussion threads for Astronomy Picture of the Day (APOD), making the resource more visible and easier to use.
\\
\\
\noindent {\em Enlist/involve others to endorse the effort}
\\
In 2011, an advisory committee was formed that includes several people prominent in the astronomical software community, computer science professors, well-known code authors, and the founders of the ASCL; among these are people who had been involved in other similar efforts. Advisors not only lend credibility and provide domain expertise, but also serve as change champions within the community. The Advisory Committee currently has ten members and has plans to eventually grow to twelve to provide better international representation.
\\
\\
\noindent {\em Provide incentives for making codes available}
\\
The ASCL started assigning unique identifiers to codes in 2011, providing a way to cite software even when it does not have a descriptive paper associated with it. The SAO/NASA Astrophysics Data System (ADS), the primary indexing service for astronomy, started indexing ASCL entries in January 2012, which makes codes in the ASCL more discoverable and also provides a way to track citations to ASCL entries. 
\\
\\
\noindent {\em Market the effort effectively}
\\
Members of the Advisory Committee developed a marketing plan that includes presenting the ASCL at conferences, writing guest pieces for blogs such as AstroBetter and Astronomy Computing Today, and using social media. Further, the APOD website, which most astronomers view at least occasionally, provides consistent and effective exposure for the ASCL by providing a link to it on the bottom of the APOD page every four to eight weeks. The editors also write to every author whose software is registered when an editor-initiated record is added.
\\
\\
\noindent {\em Engage the community} 
\\
The ASCL sponsors Special Sessions at American Astronomical Society (AAS) meetings and Birds of a Feather sessions at ADASS conferences that split the allotted time between presentations and open discussion with attendees. These sessions have been invaluable to learn what barriers to code sharing exist and what incentives may be useful to encourage sharing. These sessions also lets the ASCL learn what the community expectations are, which helps us to meet or manage these expectations. Journal editors and publishers are also part of the community, and the ASCL has engaged them in informal discussions and also through a Software Publishing Special Interest Group\footnote{http://ascl.net/wordpress/?p=1100}, and continues to do so.
\\
\\
\noindent {\em Mitigate barriers and nurture incentives}
\\
Because authors are often reluctant to deposit their software, we dropped the requirement that it be housed on the resource itself. Instead, code records have a link to the software rather than store and serve the actual source codes, as several other astronomy software directories have done. (The ASCL can and does house software, but the majority of codes it has registered reside elsewhere.) Rather than rely on scientists, who are generally very busy, to provide information on their software, we seek codes and add information about them to the ASCL. In addition to aiding citations and citation tracking for software, the ASCL advocates for greater recognition for its authors.
\\
\\
\noindent {\em Examine and reach out to other communities}
\\
As there were lessons in other astronomy efforts to build a code registry or repository, so too are there lessons from other disciplines and people who have done or are trying to do the same. Many of the challenges the ASCL faces are not unique, and with funding agencies, governments, and organizations pushing for more openness in research, opportunities to meet and discuss both small and broader issues with representatives from other disciplines periodically arise. Members of the ASCL advisory committee have participated not only in WSSSPE, but also in workshops at the Library of Congress and National Academies of Science, and have had informal discussions with representatives of fields as diverse as economics, computational chemistry, and materials science. The ASCL follows developments elsewhere, such as those discussed in Force11\footnote{https://www.force11.com}, and is a signatory to the Center for Open Science's\footnote{http://centerforopenscience.org/} Transparency and Openness Promotion (TOP) Guidelines\footnote{http://centerforopenscience.org/top/}.
\\
\\
\noindent {\em Be patient}
\\
It takes time to change a community. Our timeline is ten years; ten years after the restart of the ASCL, we should have a good idea whether the resource will reach a tipping point and become embedded into the way the community works. We are heartened by the signs of progress we are already seeing; the number of visits to the ASCL has risen every year, code authors are requesting that the ASCL entries for their software be used for citation, citations to these entries have grown substantially, and code submission by authors has increased. 

\section{And So We Grow}

We continue to expand and improve the ASCL. As the ASCL grew from 2010 from about 40 code entries to nearly 900 in mid-2014, we desired greater flexibility for managing it. A new infrastructure based on a MySQL database was proposed, developed, and went into production in July 2014. We worked with Alberto Accomazzi and Carolyn Stern Grant from ADS to normalize software author names and improve procedures for the flow of data from the ASCL to ADS. Several features were added, including automatic bibcode generation for ADS ingestion, an improved submission process, one-click author searching, and more flexible browsing. 

PHP and mySQL were chosen for their ease of use and portability. Both are powerful and ubiquitous tools for web applications. Codeigniter by EllisLab is an open source PHP framework and was chosen because it enables rapid development and increased security for the simple but fully customized user and administrative experience required for the ASCL. The framework is also fully documented\footnote{http://ellislab.com/codeigniter/user-guide/}, increasing the sustainability of the ASCL itself.

Previously, the ASCL WordPress blog and the code records were disconnected and part of two seemingly different websites. WordPress is now fully integrated into the website both as a blog and for its use as a simple content management system. The discussion forum the ASCL had been using was moved and incorporated into the new website and topic management for codes is fully automated. With the code records, blog, and forum housed under a single continuous format, user experience is now vastly less confusing and more respectable.

The new database infrastructure offers opportunities for collaboration; we hope the capabilities and flexibility of the new infrastructure can be leveraged to further increase software discovery and citation, elevate the status of those who create the programs that enable so much research, and help to make research more transparent and reproducible. 

\section{Impacts and Use}

Before the new infrastructure was implemented, the number of codes submitted by author directly to the ASCL (rather than through email) was very low, with two submitted in 2011, eight in 2012, nine in 2013, and four in the first few months of 2014, for a total of 23. Since implementation of the new site, over 100 codes have been submitted directly by authors. The sharp increase in author submissions is likely primarily due to the easier submissions process on the new site, though growing awareness within the community surely contributes as well. 

Both former and new infrastructures track how many times a record is viewed, and we see from this tracking that code records on ASCL 2.0 have been viewed over a million times (1,022,102) since mid-2010; records on the new site have been viewed over 390,000 times in less than 14 months. According to Google Analytics, overall use of the site in 2014 increased by 12\% over 2013.

ASCL IDs are used in some papers served electronically to provide a hotlink from the article directly to the code record, making it very easy to find the software used in the research discussed. An example from an article in {\em Astronomy \& Computing} \citep{2015A&C....10...22B} is shown below.

\begin{figure}[h!]
\centering
 \includegraphics[width=0.85\textwidth]{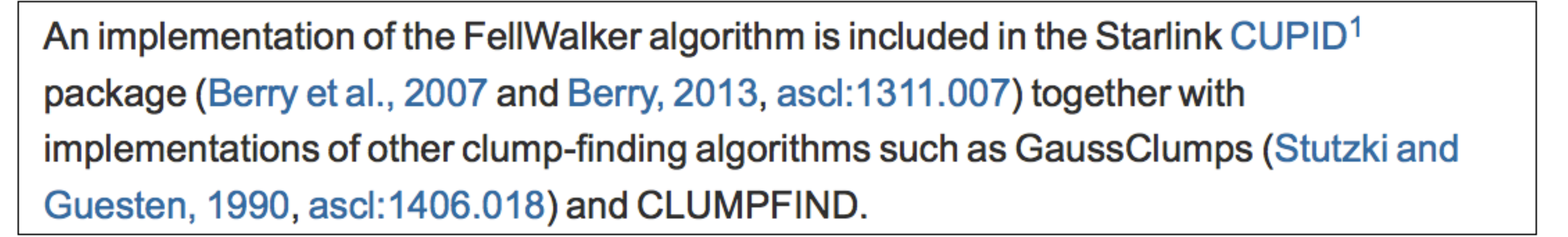}
 \caption{Screenshot showing hotlinked ASCL ids in article text}
\end{figure}

Software is cited in various ways; though the most common is citing a paper in which the software is explained or used, some authors are citing codes explicitly and independent of a code paper using ASCL entries. We started tracking the number of codes cited by their ASCL entries in January, 2014 and found that 7.5\% of the ASCL records indexed by ADS had been cited. As of the beginning of August 2015, 16.8\% of ASCL entries have been cited. The number of software records grew 34\% during this time, while the number of entries with citations grew by 70\%. 

\begin{figure}[h!]
\centering
 \includegraphics[width=0.23\textheight]{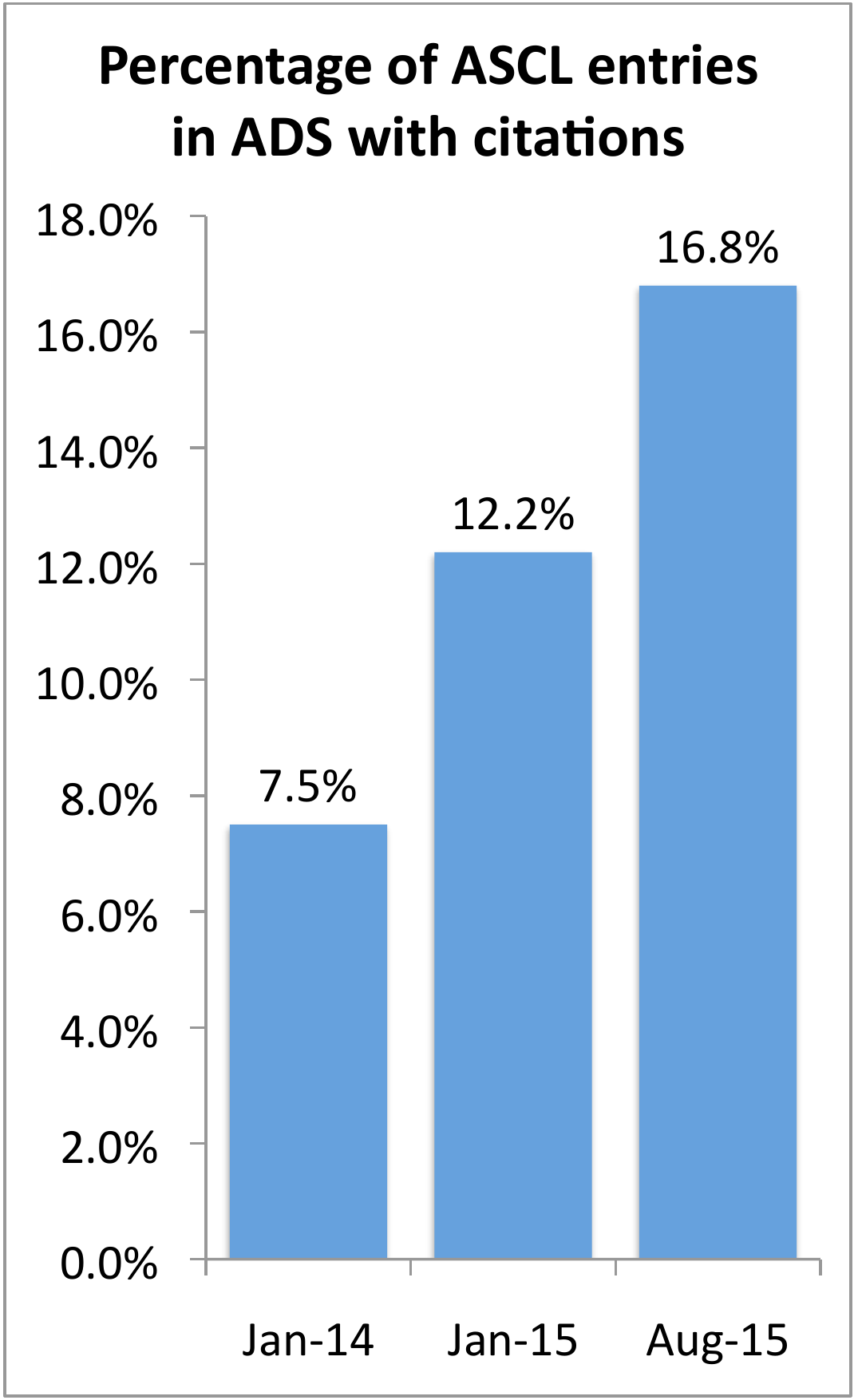}
\caption{Growth in citations to ASCL records}
\end{figure}

Code hosting sites such as GitHub and BitBucket and archival services such as Figshare and Zenodo make code available to researchers and, in the case of archival services, mint DOIs for deposited codes, are valuable resources to the community and aid in transparency of research and software sustainability. The commit IDs and DOIs available from them are sometimes used to cite software, thereby assigning credit to those who develop these tools, but at this time, ADS, the main indexing service for astrophysics, cannot track these citations; ADS can track citations only to resources it has ingested, such as ASCL entries. 

\section{Looking Forward}

Since 2010, the ASCL has taken an active approach to registering codes; its editors search the literature to discover software that enables research and creates entries for those packages that are publicly available, after which the editors inform authors of their code's entry. This has enabled the ASCL to grow much faster than it would have if relying on author submissions for the past five years, but is time-consuming. The increasing popularity of code hosting sites such as BitBucket, SourceForge, and GitHub has made discovery of new codes somewhat easier; still, approximately 85\% of codes reside elsewhere, typically on institutional or personal websites. With ASCL entries being cited and some authors and journals recognizing the benefits of a curated code registry, and with the easier submissions process, the increase in software entry by authors has shifted some of the work of ASCL editors to vetting author entries rather than discovery work. Among the improvements we'd like for the ASCL are automated methods for code discoverability for literature, code repositories, and archives, better linking between articles about and using software and the code records, and a way to represent mutable author lists for codes under active development. 

In April 2015, ASCL, ADS, GitHub, Zenodo, and other entities participated in a Sloan-sponsored meeting at GitHub in San Francisco to develop a collaboration on a cohesive research software citation-enabling platform\footnote{http://astronomy-software-index.github.io/2015-workshop/}, with the expectation that this work will improve code discoverability and transparency as well as software citation, and as of this writing, a proposal to fund this work is under development.

\section{Conclusion}

The ASCL's requirement for research software sustainability is very basic: the software must be discoverable and available for examination. Ensuring even this basic requirement can be daunting, as there are barriers to and a lack of incentives for code sharing. Looking at several attempts to create astronomy software directories with an eye toward change management suggests specific steps can be taken to provide an environment that encourages software sharing and discovery. Some of these steps are outlined, and how the ASCL has implemented them is shared with the hope that other disciplines facing some of the same challenges may find this information useful. Finally, we have provided information on recent changes and upgrades to the ASCL, how the ASCL is used, a few of the additional improvements we would like to make, and collaborative efforts underway.

\section{Acknowledgements}
The authors thank the referees, Carl Boettiger and Anonymous, for their valuable, kind, and specific feedback for improving this article; we feel their input made our article much stronger than it would have otherwise been. We thank Lior Shamir for straightening out the LaTeX mess in the first draft. The first author thanks the National Science Foundation and the Gordon and Betty Moore Foundation for travel support to attend WSSSPE2. We also thank the ASCL Advisory Committee members for their continuing guidance, Robert Nemiroff and John Wallin for letting us take the keys to the ASCL bus and drive off in it, and especially research software authors everywhere (particularly in astrophysics) for developing valuable tools for the research community even though they are not always recognized for the value they create; thank you for sharing!

\bibliographystyle{plain}
\bibliography{WSSSPEpublicationfinal}

\end{document}